\documentclass[final,1p,times]{elsarticle}
\usepackage{graphicx}
\usepackage{amssymb}
\begin{document}
\begin{frontmatter}
%
%
%
%
%
\title{The QCD Critical Point : marching towards continuum}
%
%

\author{Saumen Datta}
\ead{saumen@theory.tifr.res.in}

\author{Rajiv V.\ Gavai\corref{cor1}}
\ead{gavai@tifr.res.in}
\cortext[cor1]{Speaker at Quark Matter 2012.}

\author{Sourendu Gupta}
\ead{sgupta@theory.tifr.res.in}

\address{Department of Theoretical Physics, Tata Institute of Fundamental
         Research,\\ Homi Bhabha Road, Mumbai 400005, India.}

\begin{abstract}
We present results of our simulations of QCD with two light dynamical quarks on
a $32^3 \times 8$ lattice at a current quark mass tuned to have the Goldstone
pion mass of about 230 MeV.  Employing the Taylor expansion method we proposed
earlier, we estimate the radius of convergence of the series for the baryonic
susceptibility by using terms up to eighth order.  Together with our earlier
results, corresponding to the same physical parameters but on coarser lattices
at respectively 1.33 times and twice the lattice cut-off ($a$), we were able to
attempt a march towards the continuum limit.

\end{abstract}

\end{frontmatter}

\section{Introduction}
\label{int}

Whether the phase diagram of strongly interacting matter, governed by Quantum
Chromo Dynamics (QCD), has a critical point in the temperature ($T$) and baryon
chemical potential $\mu_B$ plane, is an exciting question that has attracted
many theorists, phenomenologists, and heavy ion experimentalists.  A variety of
models, which have been successfully tested for hadronic properties in our
world, such as an effective chiral Nambu-Jana Lasinio type model, lead to a
phase diagram \cite{WilRaj} with a critical point in a world with two light
quarks and one heavier quark.  It is clearly desirable to obtain it from QCD
directly, or show a lack of its existence, especially since enormous efforts at
RHIC and other accelerators are being devoted to look for it. One usually has
to deal with large coupling constants in the world of (low energy) hadronic
interactions.  Non-perturbative lattice QCD, defined on a discrete space-time
lattice, has proved itself to be the most reliable technique for extracting
such information from QCD.  The hadron spectrum has been computed successfully
and predictions of weak decay constants of heavy mesons have been made.  Since
these fix quark masses and $\Lambda_{\scriptscriptstyle QCD}$, a completely
parameter free application of this approach to finite temperature QCD has
yielded a slew of thermodynamics determinations, such as the pressure as a
function of temperature.  It is therefore natural to ask whether lattice QCD
can help us in locating the QCD critical point.  

Due to the well-known fermion doubling problem, one has to make a compromise in
choosing the quark type for any computation.  We employ the staggered quarks.
These have an exact chiral symmetry which provides an order parameter for the
entire $T$-$\mu_B$ plane but unfortunately flavour and spin symmetry are broken
for them on lattice.  The existence of the critical point, on the other hand,
is expected to depend crucially on the number of flavours.  Although
computationally much more expensive, Domain Wall or Overlap Fermions are better
in this regard, as they do have the correct symmetries for any lattice spacing
at zero temperature and density.  Introduction of chemical potential, $\mu$,
for these, however, is not straight-forward due to their non-locality.  Bloch
and Wettig \cite{BlWe} proposed a way to do this.  Unfortunately, it turns out
\cite{BGS} that their prescription breaks chiral symmetry .  Recently, this
problem has been solved \cite{GS12} and a lattice action with nonzero $\mu$ and
the same chiral symmetries as continuum QCD has been proposed.  It will be
interesting to compare its results with those for the staggered quarks.

Finite density simulations needed for locating a critical point suffer from
another well known problem.  This one is inherited from the continuum theory
itself: the fermion sign problem. Assuming $N_f$ flavours of quarks, and
denoting by $\mu_f$ the corresponding chemical potentials, the QCD partition
function is 

\begin{equation}
{\cal Z} =  \int D U \exp(-S_G) ~ \prod_{f} {\rm
Det}~M(m_f, \mu_f)~~, 
\end{equation}   
where quarks have been integrated leading to the determinant. The thermal 
expectation value of an observable ${\cal O}$ is
\begin{equation}
 \langle {\cal O} \rangle = \frac{   \int D U \exp(-S_G) ~ {\cal O} \prod_{f} {\rm Det}~M(m_f, \mu_f)} { {\cal Z~~}} .  
\end{equation}   

Numerical simulations or analytical computations can be done if Det~$M > 0$ for
any set of $\{U\}$.  However, Det~$M$ is a complex number for any $\mu \ne 0$.
This famous phase/sign problem is a major stumbling block in extending the
lattice techniques to the entire $T$-$\mu_B$ plane.  Several approaches have
been proposed in the past decade to deal with it.  Let me provide a partial
list: 1) Two parameter Re-weighting \cite{FoKa}, 2) Imaginary Chemical Potential \cite{ImMu}, 3) Taylor Expansion \cite{TaEx}, 4) Canonical Ensemble 
method \cite{CaEn}, and 5) Complex Langevin approach \cite{ComLa}.
We employ the Taylor expansion approach \cite{TaEx} to obtain 
the results discussed in the next section.

\section{Lattice Results}
\label{olr}

Our earlier results were obtained by simulating full QCD with two flavours of
staggered fermions of mass $m/T_c =0.1$ on $N_t \times N_s^3$ lattices, with
$N_t=4$ and $N_s=$ 8, 10, 12, 16, 24 \cite{our1} and a finer $N_t= 6 $ with
$N_s=$ 12, 18, 24 \cite{our2}. From the work of the MILC collaboration
\cite{milc}, we know that our lattices correspond to $m_\pi/m_\rho = 0.31 \pm
0.01$, leading to a Goldstone pion of 230 MeV.  In order to approach the
continuum limit of $a \to 0$,  we proceeded to work on an even finer lattice $8
\times 32^3$, keeping  $m/T_c =0.1$ fixed.  As before, the
peak of the Polyakov loop susceptibility was used to define the critical
coupling $\beta_c$ and plaquette determinations on symmetric lattices were used
to tune the temperature such that simulation range covered $0.90 \le T/T_c \le
2.01$.  Our determination of $\beta_c$ is in excellent agreement with that of
Gottlieb et al. \cite{gott8}.  Typically 100-200 independent configurations,
separated by many autocorrelation lengths were used to make measurements of 
physical quantities.

From canonical definitions of number densities $n_i$ and 
susceptibilities   $\chi_{n_u,n_d}$, the QCD pressure $P$ can be seen to have
the following expansion in $\mu$:
\begin{equation}
   \frac{\Delta P}{T^4} \equiv \frac{P(\mu, T)}{T^4} - \frac{P(0, T)}{T^4}
   = \sum_{n_u,n_d} \chi_{n_u,n_d}\;
        \frac{1}{n_u!}\, \left( \frac{\mu_u}{T} \right)^{n_u}\, 
        \frac{1}{n_d!}\, \left( \frac{\mu_d}{T} \right)^{n_d}\, 
\end{equation} 
where the indices $n_u$ and $n_d$ denote the number of derivatives of the 
partition function with respect to the corresponding chemical potentials.
We set $\mu_u=\mu_d = \mu_B/3$ and $m_u = m_d$ in the expressions and
construct a series for baryonic susceptibility from this expansion \cite{our1}.
Its radius of convergence is what we look for. 

Successive estimates for the radius of convergence were obtained by
using the ratio method [$r_{n+1/n+3} = \sqrt{{n(n+1)\chi^{(n+1)}_B}/
{\chi^{(n+3)}_B T^2}}$ ] and the n$^{th}$ root method [$ r_{2/n} = 
(n!{\chi_B^{(2)}}/{\chi_B^{(n+2)} T^n} )^{1/n}$].  We used
terms up to 8th order in $\mu$ for doing so.  A key point to note is
that all coefficients of the series must be positive for the critical point to
be at real $\mu$, and thus physical. We thus first look for this condition to
be satisfied and then look for agreement between the two definitions above as
well as their $n$-independence to locate the critical point. The detailed
expressions for all the terms can be found in \cite{our1} where the method to
evaluate them is also explained.  We use stochastic estimators. For terms up to
the 8$^{th}$ order one needs 20 inversions of $M$ on $\sim$1000 vectors for a
single measurement on a given gauge configuration.  Work is in progress to
double these.  Our earlier determination of the critical point on $N_t
= 6$ resulted from the constancy for both the ratios defined above at $T/T_c
=0.94$, with all the susceptibilities being positive, leading \cite{our2} to
the coordinates of the endpoint (E) --the critical point--to be $ T^E/T_c =
0.94\pm0.01$, and $\mu_B^E/T^E = 1.8\pm0.1$ for that lattice.

\begin{figure}[htb]
\includegraphics[scale=0.5]{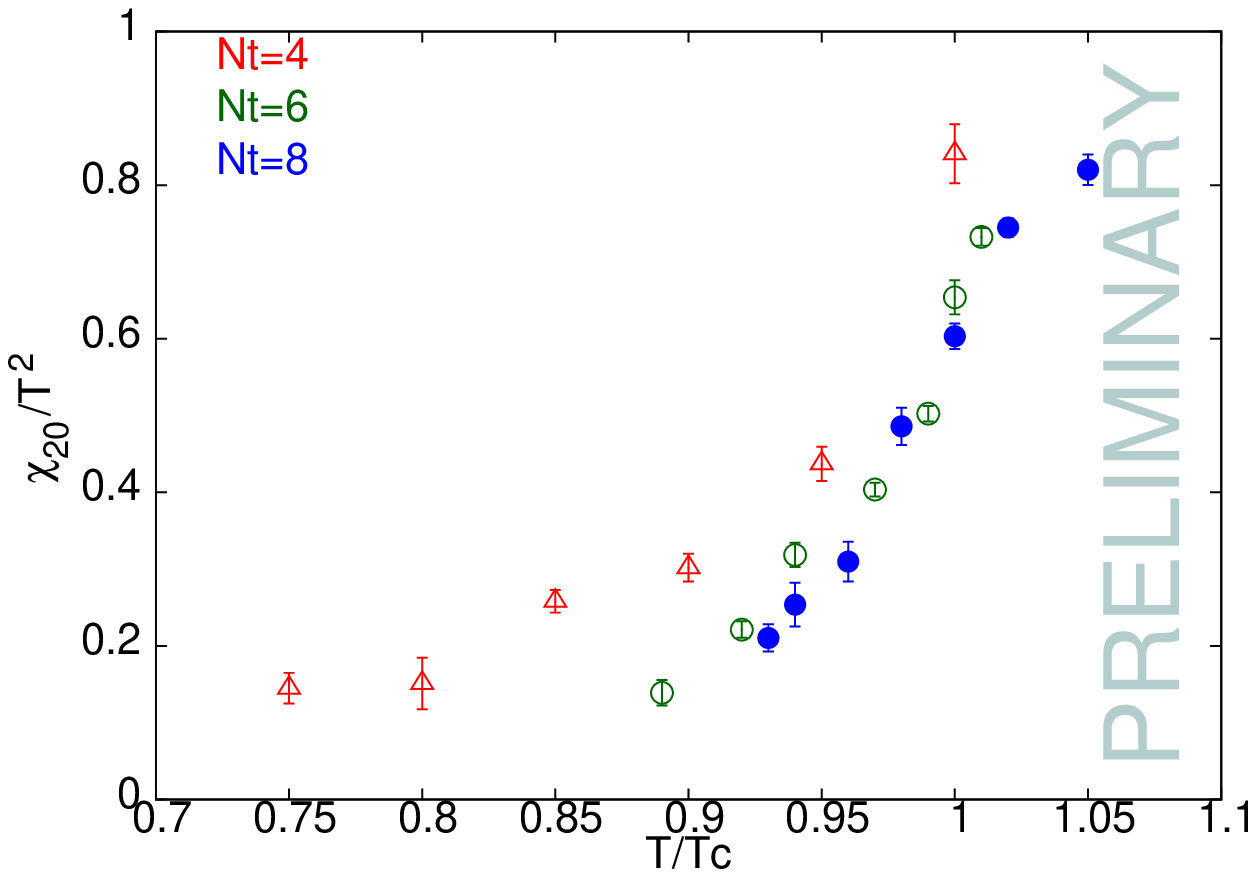}
\includegraphics[scale=0.5]{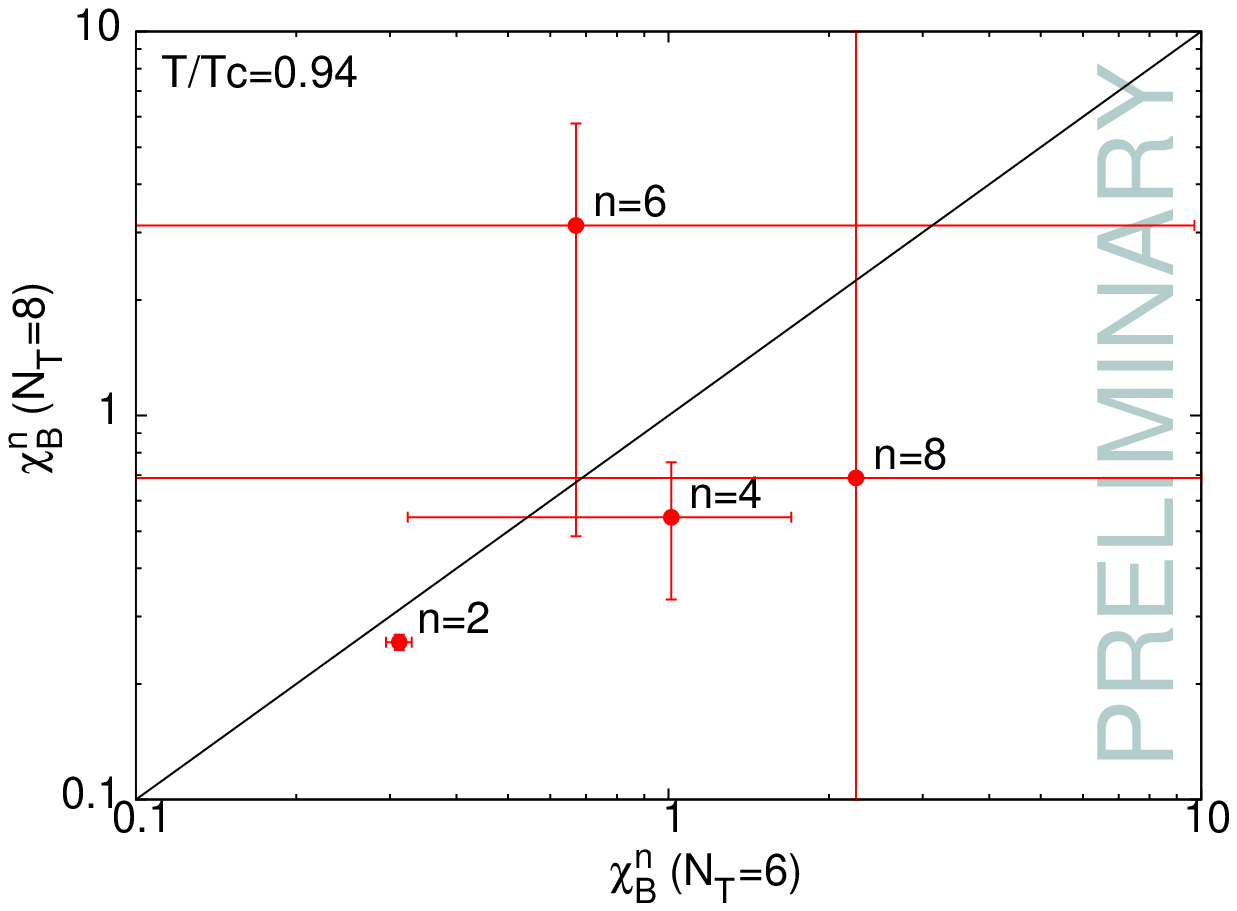}
\caption{ Comparison of baryon number susceptibility on 
$N_t$ = 8 with our earlier results on $N_t$ =6 and 4 (Left panel).  
Comparison of the expansion coefficients on $N_t=6$ and $8$ lattices.
The diagonal solid line displays the ideal case with no cut-off 
effects.  (Right).  }
\label{rads}
\end{figure}

The left panel of Fig. \ref{rads} shows a comparison of our new results for the
$N_t= 8$ lattice for baryon number susceptibility with those for $N_t=6$ and 4.
The encouraging agreement between $N_t = 8$ and 6 suggests that the
dimensionless ratios which we employ in all our critical point determinations
possess only a mild cut-off dependence.  The right panel extends this
comparison to all the susceptibilities we determined for the two lattices with
$N_t=8$ and 6.  The diagonal line indicates the trajectory of their expected
locations in the ideal case of no lattice cut-off effects.  This is seen to be
so within errors, which still need to be reduced further.

\begin{figure}[htb]
\includegraphics[scale=0.5]{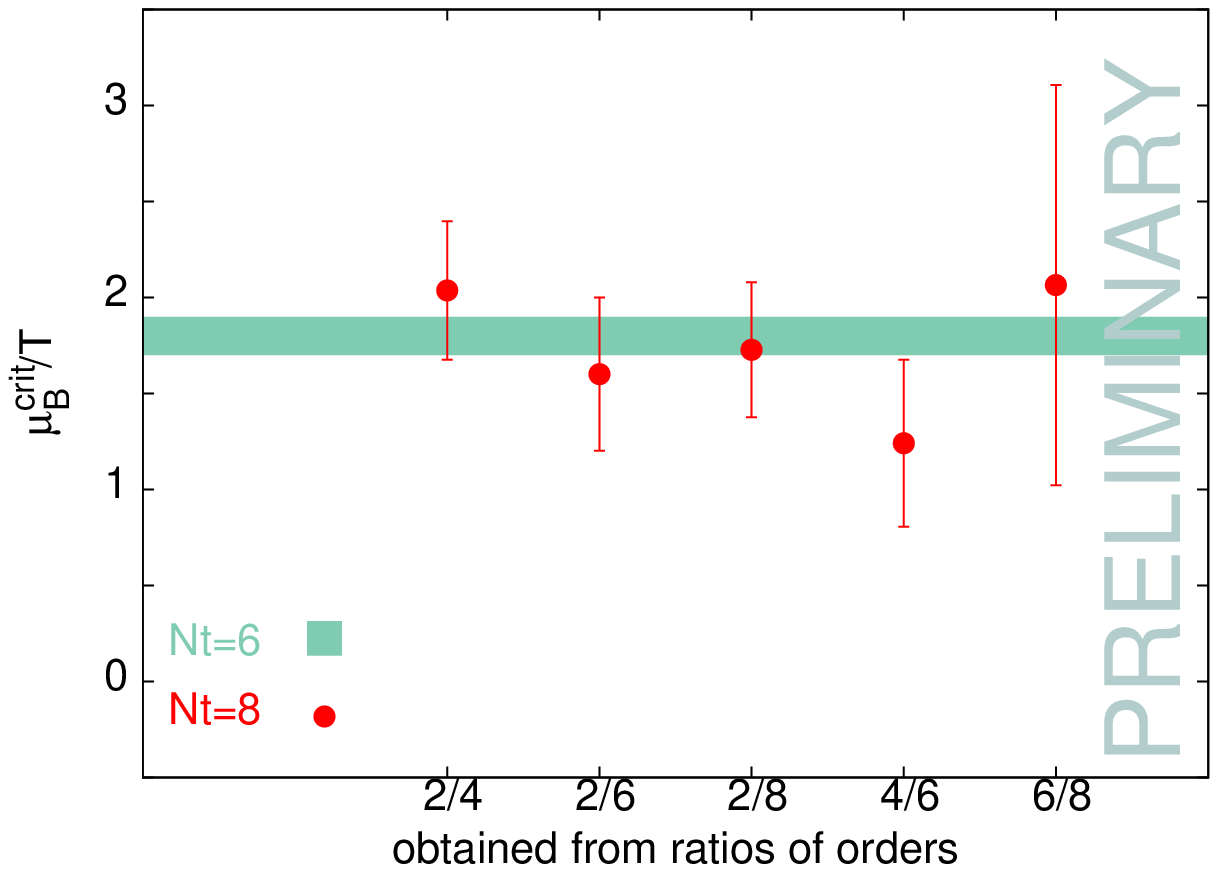}
\includegraphics[scale=0.5]{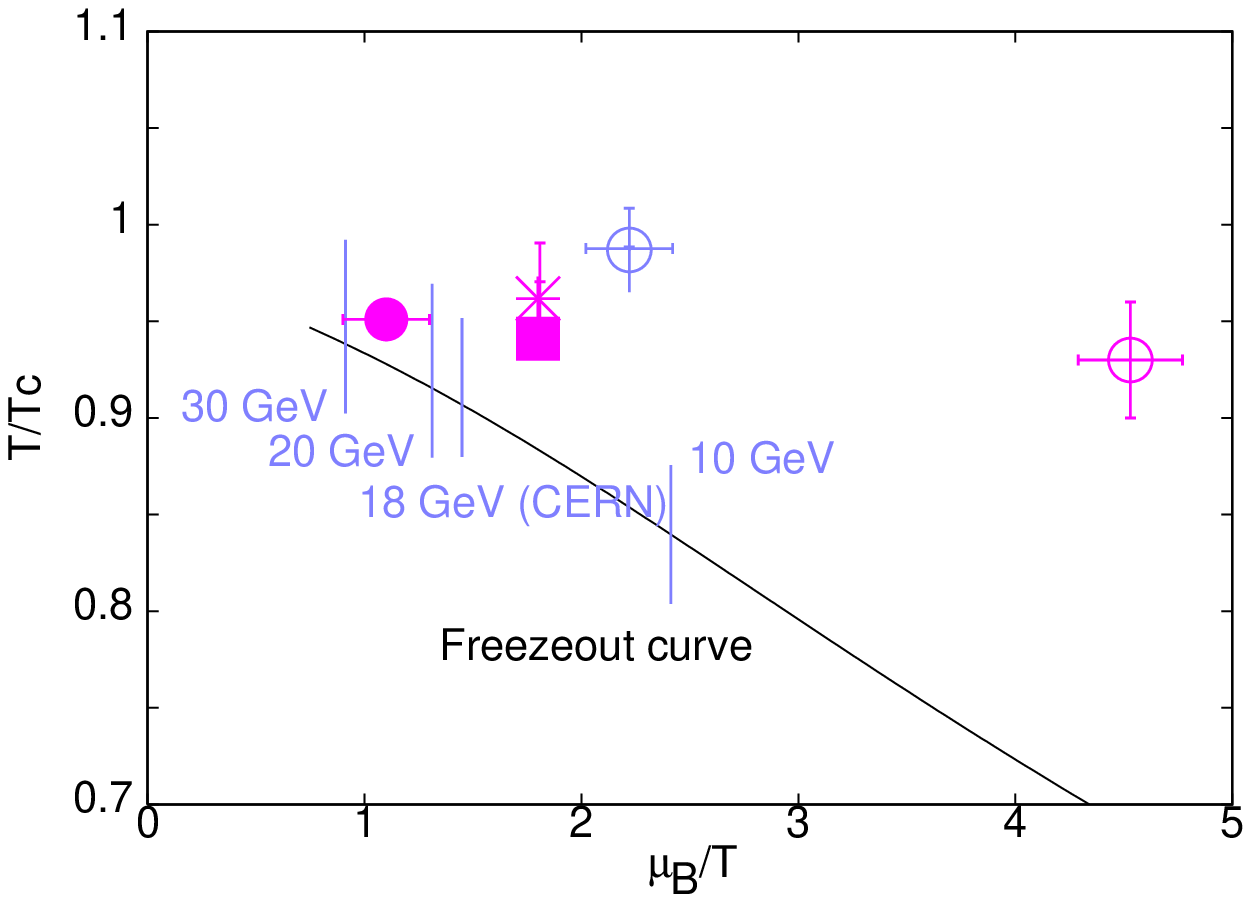}
\caption{ The radius of convergence estimates for $N_t =8$ (data) along with
the $N_t =6$ critical chemical potential (band) at the critical temperature
of $T_E/T_c = 0.94$ (Left panel).  QCD phase diagram with all known
lattice determinations for critical point (Right). Our new $N_t =8$ results,
presented in this talk, are denoted by the asterisk.  $N_t=4$ \cite{FoKa,our1} 
and 6 \cite{our2} results are shown by the circles and square respectively.  }
\label{polypad}
\end{figure}

Fig. \ref{polypad} in its left panel displays our new estimates
for the $N_t = 8$, obtained by using the two methods mentioned above and the
coefficients in Fig. \ref{rads}.  The solid band indicates our critical chemical
potential estimate on $N_t =6$ and the temperature chosen is the same as the
corresponding critical temperature $T_E$.  We see such behaviour in a small
band near this temperature, leading to a larger error band on $T_E$, as
exhibited in the QCD phase diagram in the right panel along with our old
results for $N_t=6$ \cite{our2} and 4 \cite{our1}, and those from
Budapest-Wuppertal group both \cite{FoKa} of which use $N_t =4$.

\section{Summary}
\label{su}

The elusive QCD phase diagram in $T$-$\mu_B$ plane has begun to emerge using
first principles lattice approach.  Our lattice results for $N_t =8$ are 
in very good agreement with those for $N_t =6 $, suggesting the continuum
limit to be in sight and the critical point estimate to be robust.
 
This work was done on the Blue Gene P of Indian Lattice Gauge Theory 
Initiative, Tata Institute (TIFR), Mumbai. We gratefully acknowledge 
financial and technical support of TIFR.

\end{document}